\documentclass[a4paper,11pt]{article}
\usepackage{pos}
\usepackage{amsmath}
\usepackage{braket}
\usepackage{macros_AS} 

\title{Gravitational form factors of glueballs in Yang-Mills theory}
\ShortTitle{}

\author[a]{Ryan Abbott}
\author[b]{Daniel C. Hackett}
\author*[c,d]{Dimitra A. Pefkou}
\author[e,a]{Fernando~Romero-López}
\author[a]{Phiala Shanahan}

\affiliation[a]
{Center for Theoretical Physics, Massachusetts Institute of Technology, Cambridge, MA 02139, USA}
\affiliation[b]
{Fermi National Accelerator Laboratory, Batavia, IL 60510, USA}
\affiliation[c]{Department of Physics, University of California, Berkeley, CA 94720, USA}
\affiliation[d]{Nuclear Science Division, Lawrence Berkeley National Laboratory, Berkeley, CA 94720, USA}
\affiliation[e]
{Albert Einstein Center, Institute for Theoretical Physics, University of Bern, 3012 Bern, Switzerland}

\emailAdd{dpefkou@berkeley.edu}

\abstract{
This work presents preliminary results of the first determination of the energy-momentum tensor form factors of the scalar glueball, referred to as gravitational form factors (GFFs). The calculation has been carried out in lattice Yang-Mills theory at a single lattice spacing. Using variationally optimized operators, the matrix elements are extracted from ratios of three-point functions to two-point functions. The glueball GFFs and their kinematic dependence are compared to those of other hadrons from previous calculations.
}

\FullConference{The 41st International Symposium on Lattice Field Theory (LATTICE2024)\\
 28 July - 3 August 2024\\
Liverpool, UK \\
Preprint numbers: MIT-CTP/5787, FERMILAB-CONF-24-0635-T }

\begin{document}
\maketitle

\section{Introduction}

The potential existence of glueballs---hadronic states with purely gluonic degrees of freedom---has been speculated since the inception of quantum chromodynamics (QCD)~\cite{Fritzsch:1972jv}. Nowadays, there are multiple experimental candidates for various glueball states with allowed quantum numbers $J^{PC} = 0^{++},0^{-+},1^{++},1^{+-},1^{--},2^{++}$, etc.~\cite{Klempt:2007cp,Crede:2008vw,Chen:2022asf,BESIII:2010gmv,BESIII:2019wkp,BESIII:2023wfi}. Lattice QCD calculations of the glueball spectrum have provided indispensable inputs to this search~\cite{Athenodorou:2020ani}---see Ref.~\cite{Vadacchino:2023vnc} for a review. 
However, conclusive identification of observed hadrons as glueballs or glueball-like remains challenging, calling for further theory predictions of the properties of these states to compare against.
 
Information about the internal structure of hadrons may allow classification of observed hadron states as glueball-like objects. Features like their radius, or the momentum fraction carried by gluons, could serve as smoking-gun evidence of a hadron having predominantly gluonic degrees of freedom. Both of these quantities, as well as additional information like the energy distribution, are contained in their \textit{gravitational form factors} (GFFs), which are defined from the matrix elements of the \textit{energy-momentum tensor} (EMT) $T^{\mu\nu}$ of QCD~\cite{Lorce:2018egm,Polyakov:2002yz,Polyakov:2018zvc,Burkert:2023wzr}. These matrix elements, and consequently the glueball GFFs, can in principle be determined using lattice QCD.

In this work, we take a step in this direction by studying the GFFs of the scalar glueball $G[0^{++}]$ in $SU(3)$ Yang-Mills theory.\footnote{There is one previous attempt in literature to obtain form factors of a $G[0^{++}]$ state, using an $SU(2)$ pure gauge action and a plaquette as the probe~\cite{Tickle:1989gw}.} With a pure gauge action, the EMT contains only a gluonic contribution
${T^{\mu\nu}= 2\;\text{Tr}[-F^{\mu}_{\alpha}F^{\alpha\nu}+\frac{1}{4}g^{\mu\nu}F^{\alpha\beta}F_{\alpha\beta}]}$,
where $F^{\mu\nu}$ is the gluon strength tensor. A scalar glueball has two GFFs, $A(t)$ and $D(t)$, defined 
in the EMT matrix element decomposition as
\begin{equation} \label{eq:ME}
    \left\langle G[0^{++}](p') \right| T^{\mu\nu} \left| G[0^{++}](p)\right\rangle = 2 P^{\mu}P^{\nu}A(t)+\frac{\Delta^{\mu}\Delta^{\nu}-g^{\mu\nu}\Delta^2}{2}D(t) \;,
\end{equation}
where $p$ and $p'$ are the four-momenta of the incoming and outgoing states,
$P=(p+p')/2$, $\Delta=p'-p$, and $t=\Delta^2$. The momentum sum rule dictates that $A(0)=1$, while $D(0)$, also known as the $D$-term, is unconstrained.

\section{Lattice setup}

The results in this work are obtained from a single lattice ensemble on volume $L^3 \times T = 24^3\times 48$ 
for the purely gluonic theory defined with the $SU(3)$ Wilson gauge action with $\beta=5.95$. 
Setting the scale with the Sommer parameter gives $a=0.098$ fm~\cite{Necco:2001xg,Durr:2006ky}.
We generate $\mathcal{O}(10^7)$ 
configurations using $\mathcal{O}(10^5)$ independent streams of heatbath and overrelaxation~\cite{Creutz:1980zw, Cabibbo:1982zn, Kennedy:1985nu, Brown:1987rra, Adler:1987ce}. 
On each configuration, we measure correlation functions 
constructed using two interpolating operators:
\begin{equation} \label{eq:interp}
\chi_1(x) = \frac{1}{4}\sum_{\mu\neq\nu}\text{ReTr} \, U^2_{\mu\nu}(x), \quad \chi_2(x) = \frac{1}{4}\sum_{\mu\neq\nu}\text{ReTr} \, U^7_{\mu\nu}(x) \;,
\end{equation}
where $\mu,\nu\in\{x,y,z\}$, and $U^n_{\mu\nu}$ is 
an $n \times n$ Wilson loop constructed from links stout-smeared~\cite{Morningstar:2003gk} by 3 steps in spatial directions only. The absence of fermionic fields in the operators allows computing and working directly with interpolators of definite three-momentum $\vec{p}$, defined from Eq.~\eqref{eq:interp} with their vacuum expectations subtracted as
\begin{equation}
    \chi_i(\vec{p}, t) = \sum_{\vec{x}} e^{-i \vec{p} \cdot \vec{x}} 
    \left[ \chi_i(\vec{x},t) - \braket{\chi_i(\vec{x},t)} \right] .
    \label{eq:interp-p}
\end{equation}
The summations over $\mu,\nu$ in Eq.~\eqref{eq:interp} project to the $A_1^+$ (rest frame) or $A_1$ (moving frames) irreducible representation (irrep) of the finite-volume symmetry group, and taking the real part projects to positive charge conjugation quantum numbers. The lowest-energy state excited by these interpolators is the positive parity $0^{++}$ glueball. However, 
above the $0^{++}$ ground state, the spectrum may also include
glueballs with other quantum numbers, e.g.~tensor or pseudoscalar glueballs, or multi-glueball or ditorelon states, depending on the momentum frame.

We use the clover definition of $F_{\mu\nu}$,
\begin{equation}
F_{\mu\nu}(x) = \frac{i}{8 g_0}(Q_{\mu\nu}(x) - Q^{\dagger}_{\mu\nu}(x)) \;,
\end{equation}
where 
\begin{equation}
  \begin{split}
Q_{\mu\nu}(x) = &U_{\mu}(x)U_{\nu}(x+\hat{\mu})U^{\dagger}_{\mu}(x+\hat{\nu})
U^{\dagger}_{\nu}(x) \\
+ &U_{\nu}(x)U_{\mu}^{\dagger}(x-\hat{\mu}+\hat{\nu})U_{\nu}^{\dagger}(x-\hat{\mu})
U_{\mu}(x-\hat{\mu}) \\
+ &U^{\dagger}_{\mu}(x-\hat{\mu})U^{\dagger}_{\nu}(x-\hat{\mu}-\hat{\nu})
U_{\mu}(x-\hat{\mu}-\hat{\nu})U_{\nu}(x-\hat{\nu}) \\
+ &U^{\dagger}_{\nu}(x-\hat{\nu})U_{\mu}(x-\hat{\nu})U_{\nu}(x-\hat{\nu}+\hat{\mu})
U^{\dagger}_{\mu}(x) 
\end{split}  
\end{equation}
to construct the EMT, and compute it from links stout-smeared in all directions by 3 steps.
Vacuum-subtracted and projected to definite three-momentum $\vec{\Delta}$, the operators of interest are
\begin{equation}
    T_{\mathcal{R} \ell}(\vec{\Delta}, \tau) = \sum_{\vec{y}} e^{i \vec{\Delta} \cdot \vec{y}} \left[ T_{\mathcal{R} \ell}(\vec{y},\tau) - \braket{T_{\mathcal{R} \ell}(\vec{y},\tau)} \right]
\end{equation}
where $\mathcal{R} \in \{ \tau_1^{(3)}, \tau_3^{(6)} \}$ denotes the irrep of the hypercubic group that the symmetric traceless $T_{\mu\nu}$ is subduced to in Euclidean space and $\ell$ indexes the irrep bases. We use the same complete orthonormal irrep bases~\cite{Gockeler:1996mu} as in previous works on GFFs, e.g., Refs.~\cite{Brommel:2007zz,Detmold:2017oqb,Shanahan:2018pib,Shanahan:2018nnv,Hackett:2023rif,Hackett:2023nkr}, i.e.,
 \begin{equation} \label{eq:irrep3basis}
\begin{split}
T_{\tau_{1,1}^{(3)}} &=  \frac{1}{2}(T_{00}+T_{11}-T_{33}+T_{00}),\\ 
T_{\tau_{1,2}^{(3)}} = \frac{1}{\sqrt{2}}&(T_{11}-T_{22}),\; T_{\tau_{1,3}^{(3)}} = \frac{1}{\sqrt{2}}(T_{33}+T_{00})\;,
\end{split}
\end{equation}
for $\tau_1^{(3)}$ and 
\begin{equation}
\begin{split} \label{eq:irrep6basis}
T_{\tau_{3,1}^{(6)}} = \frac{1}{\sqrt{2}}(T_{12}+T_{21}),&\;T_{\tau_{3,2}^{(6)}} =\frac{1}{\sqrt{2}}(T_{13}+T_{31}) \;,\\
T_{\tau_{3,3}^{(6)}} =\frac{-i}{\sqrt{2}}(T_{10}+T_{01}),&\;T_{\tau_{3,4}^{(6)}} =\frac{1}{\sqrt{2}}(T_{23}+T_{32}) \;,\\
T_{\tau_{3,5}^{(6)}} =\frac{-i}{\sqrt{2}}(T_{20}+T_{02}),&\;T_{\tau_{3,6}^{(6)}} =\frac{-i}{\sqrt{2}}(T_{30}+T_{03}) \;,
\end{split}
\end{equation}
for $\tau_3^{(6)}$, both written in Minkowski space.

\section{Glueball spectrum}

\begin{figure}
\includegraphics[width=0.98\linewidth]{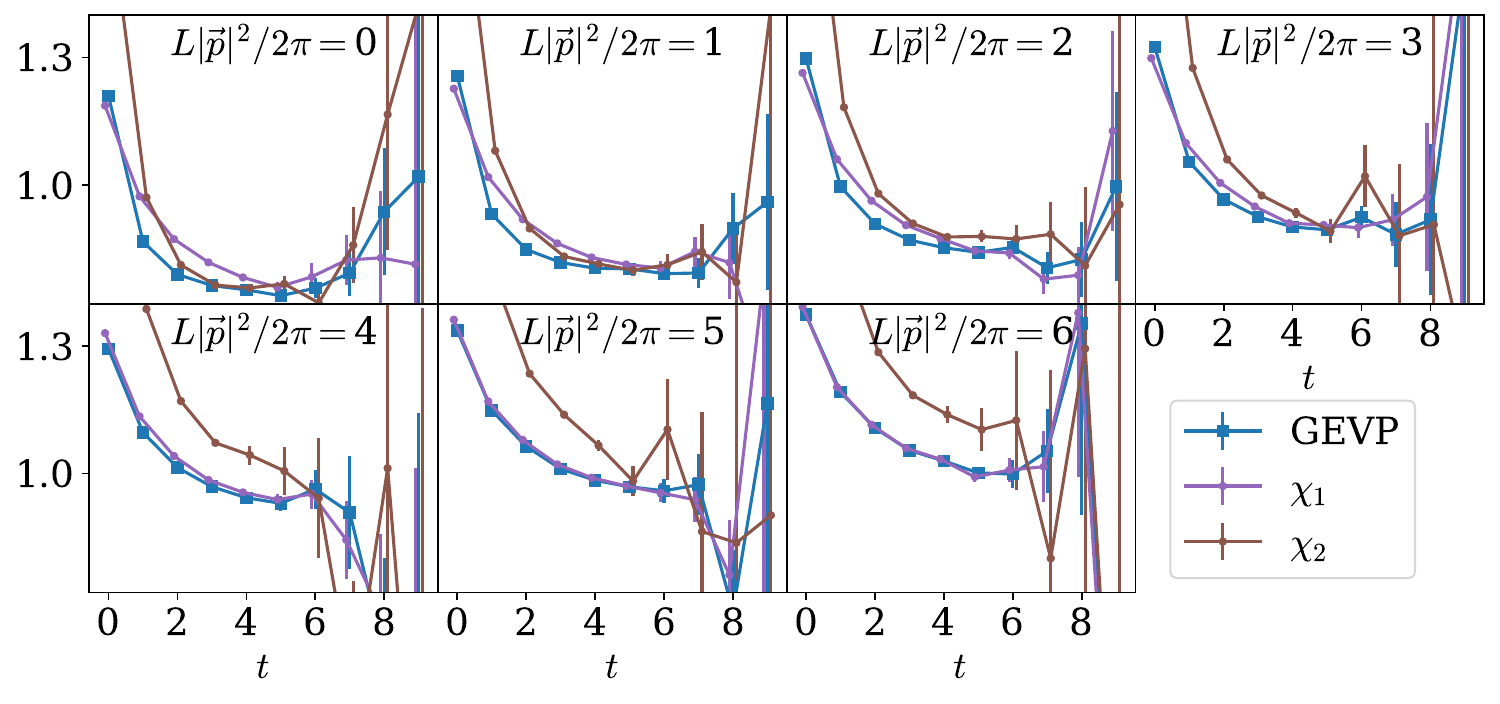}
\caption{
    Effective energies $E_{\text{eff}}(t)=\text{log}(C^{2\text{pt}}(t)/C^{2\text{pt}}(t+1))$ as a function of sink time $t$ for the different boost momenta, comparing the diagonal correlators $C^{2\text{pt}}_{ii}$ of Eq.~\eqref{eq:2pt-matrix} computed with single interpolators from the basis Eq.~\eqref{eq:interp} with the GEVP-optimized one of Eq.~\eqref{eq:gevp-2pt} defined with the composite interpolator Eq.~\eqref{eq:interp-gevp}
}
\label{fig:spectrum}
\end{figure}

The analysis begins by determining the spectrum and constructing optimized ground-state interpolating operators.
To proceed, we compute $2 \times 2$ matrices of momentum-projected, vacuum-subtracted two-point functions averaged over all timeslices
\begin{equation}
C^{2\text{pt}}_{ij}(\vec{p},t) = \frac{1}{T} \sum_{t_0} \braket{ \chi_i(\vec{p},t+t_0) \, \chi_j(\vec{p}, t_0)^\dagger }
\label{eq:2pt-matrix}
\end{equation}
for all $\left|\vec{p}\right|^2\leq 6 (2\pi/L)^2$ on 200 bootstrap ensembles after binning the $\mathcal{O}(10^7)$ configurations into groups of $1000$.
We average over equivalent momenta to obtain two-point functions for the 7 distinct $|\vec{p}|^2$.
For each $\left| \vec{p} \right|^2$, we then solve the generalized eigenvalue problem (GEVP) to extract the ground state, which we identify as the scalar glueball. 
Employed in a ``fixed pivot'' mode with $t_0=1$ and diagonalization time $t_d=3$,
the GEVP provides 7 sets of weights $w_{ij}{(|\vec{p}|^2)}$, one for each distinct $|\vec{p}|^2$, which we use to construct optimized interpolators
\begin{equation}
    \chi_{0}(\vec{p},t) = \sum_i w_{0i}{(|\vec{p}|^2)} \chi_i(\vec{p},t) .
    \label{eq:interp-gevp}
\end{equation}
From these, two-point functions are obtained as 
\begin{equation}
    C_{0^{++}}^{2\text{pt}}(\vec{p},t) 
    = \frac{1}{T} \sum_{t_0} \braket{\chi_{0}(\vec{p},t+t_0) ~ \chi_{0}(\vec{p},t_0)^\dagger }
    = \sum_{i,j} w_{0i}(|\vec{p}|^2) \, C^{2\mathrm{pt}}_{ij}(\vec{p},t) \, w^*_{0j}(|\vec{p}|^2) ,
    \label{eq:gevp-2pt}
\end{equation}
labeled by the quantum numbers of the scalar glueball.
Figure~\ref{fig:spectrum} compares the ground-state energies extracted using GEVP-optimized interpolators versus the individual interpolators in the basis.

\section{Matrix elements}

\begin{figure}
\includegraphics[width=0.98\linewidth]{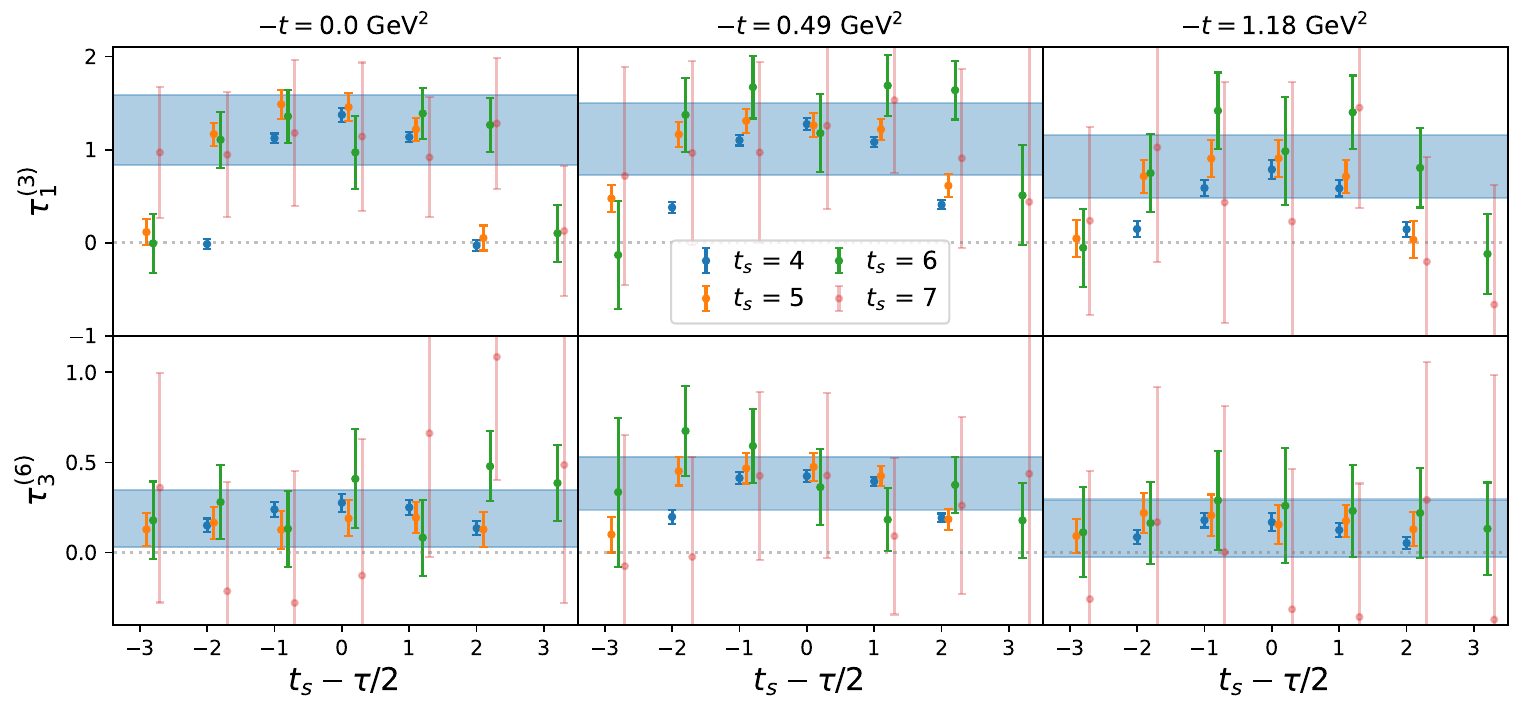}
\caption{Example ratios of three- and two-point functions at different sink times $t_s$, expected to be proportional to the matrix elements of Eq.~\eqref{eq:ME} up to known kinematic factors. Deviation from a constant value (i.e., curvature) is an indication of excited-state contamination effects. The bands correspond to model averages over fits to different time ranges. The two rows show examples from each of the two irreps, and the three columns to different momentum transfers $t$.}
\label{fig:ME}
\end{figure}

To obtain the matrix elements defined in Eq.~\eqref{eq:ME}, 
we use the GEVP-optimized interpolators $\chi_0$ to compute vacuum-subtracted three-point functions~\cite{Blossier:2009kd}
\begin{equation}\begin{aligned}
    C^{3\mathrm{pt}}_{0^{++},\mathcal{R}\ell}(\vec{p'},\vec{\Delta},t_s,\tau) 
    &= \frac{1}{T} \sum_{t_0} \braket{ \chi_0(\vec{p}', t_s+t_0) \, T_{\mathcal{R} \ell}(\vec{\Delta}, \tau+t_0) \, \chi_0(\vec{p}', t_0)^\dagger }
\end{aligned}\end{equation}
for all $\left|\vec{p}'\right|^2 \leq 6 (2\pi/L)^2$ and $\left|\vec{\Delta}\right|^2 \leq 10 (2\pi/L)^2$.
To isolate the ground-state matrix element, we form the standard ratios to cancel the leading overlap factors and time dependence
\begin{equation}
    R_{0^{++},\mathcal{R}\ell}(\vec{p'},\vec{\Delta},t_s,\tau)
    = \frac{ 
        C^{3\mathrm{pt}}_{0^{++},\mathcal{R}\ell}(\vec{p'},\vec{\Delta},t_s,\tau)
    }{ C^{2\mathrm{pt}}_{0^{++}}(\vec{p'}, t_s) }
    \sqrt{
        \frac{ C^{2\mathrm{pt}}_{0^{++}}(\vec{p}, t_s-\tau) }
             { C^{2\mathrm{pt}}_{0^{++}}(\vec{p}', t_s-\tau) }
        \frac{ C^{2\mathrm{pt}}_{0^{++}}(\vec{p}', t_s) }
             { C^{2\mathrm{pt}}_{0^{++}}(\vec{p}, t_s) }
        \frac{ C^{2\mathrm{pt}}_{0^{++}}(\vec{p}', \tau) }
             { C^{2\mathrm{pt}}_{0^{++}}(\vec{p}, \tau) }
    } . 
\end{equation}
The resulting quantities are proportional to the matrix elements of interest up to known kinematic factors and excited state effects. Thus, to extract the matrix elements, we fit a constant to all possible ($t_s$,$\tau$) time ranges with at least 8 data points within the constraints that $t_{s_{\text{min}}} \geq 5$, $t_{s_{\text{max}}} \leq 15$, $\tau \geq 2$,  $\tau \leq t_s - 2$ of each ratio and model-average over the resulting set of fits with AIC weights~\cite{Jay:2020jkz}. 
Example ratios and fits thereof are shown in Fig.~\ref{fig:ME}.

\section{Gravitational form factors}

\begin{figure} 
\includegraphics[width=0.98\linewidth]{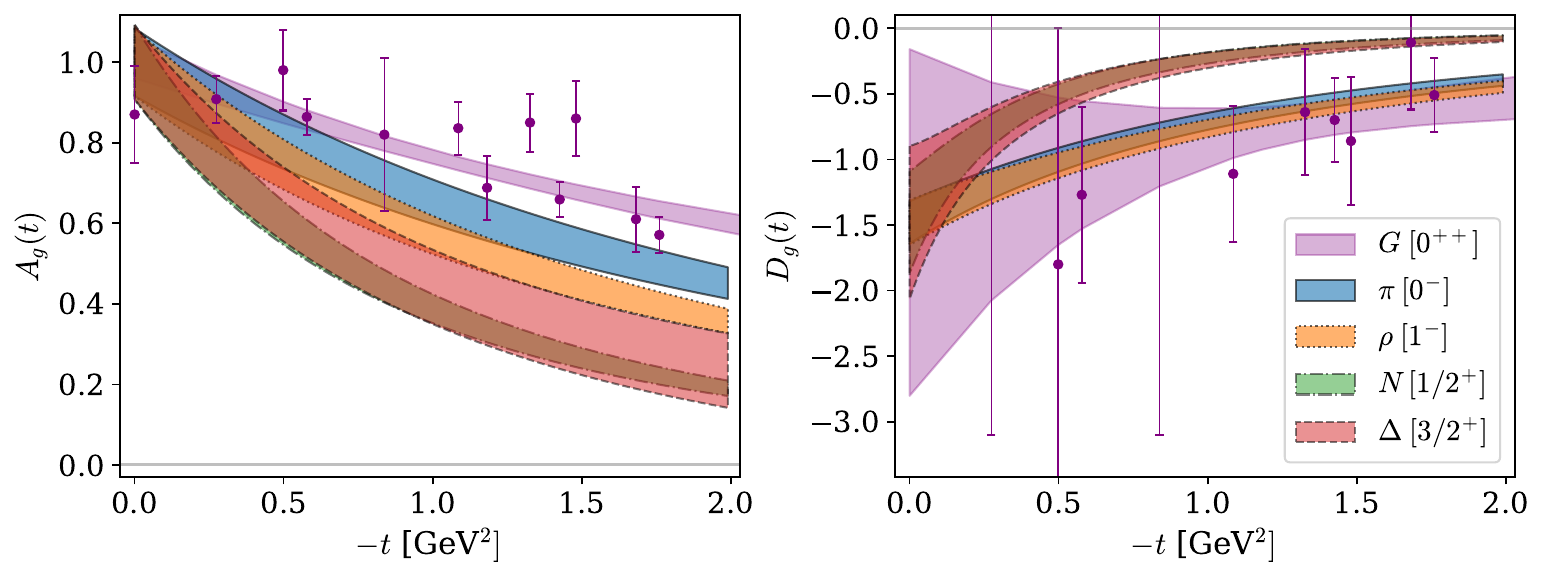}
\caption{
Comparison between the $G[0^{++}]$ glueball GFFs in Yang-Mills theory obtained in this work, and the gluon GFFs of 
four other hadrons---the pion, $\rho$ meson, nucleon, and $\Delta$ baryon, indicated with their $J^P$ quantum numbers---obtained 
with an $N_f=2+1$ QCD ensemble with $m_{\pi}=450~\text{MeV}$~\cite{Pefkou:2021fni}.  The latter have been normalized to match the values of the glueball $A(0)$ and $D(0)$. 
}
\label{fig:GFFs}
\end{figure}

The GFFs are obtained by first
grouping the data for each irrep separately into 12 
bins using k-means clustering \cite{kmeans1d} on the momentum transfer squared $t=\Delta^2$, 
then solving the overconstrained systems of linear equations dictated by Eq.~\eqref{eq:ME} to obtain the bare GFFs $A_\mathcal{R}(t)$ and $D_\mathcal{R}(t)$ for each bin and irrep. 
These
may be renormalized by imposing the sum rule $A(0)=1$.
The renormalization factors $1 / A^{\mathrm{bare}}_{\mathcal{R}}(0)$ are obtained 
from a fit of a dipole model $\alpha / (1+t/\Lambda^2)^2$
to the bare GFF $A_{\mathcal{R}}(t)$, where $\alpha$ and $\Lambda$ are fitted parameters, identifying $\alpha = A^{\mathrm{bare}}_{\mathcal{R}}(0)$.
The bare GFFs in each momentum 
bin for each irrep are then multiplied by these factors, averaged together, and fit again with dipoles to obtain the results shown in Fig.~\ref{fig:GFFs}.

Figure~\ref{fig:GFFs} compares the results for $A(t)$ and $D(t)$ of the $0^{++}$ glueball against the gluon GFFs obtained for four hadrons with quantum numbers $J^P=0^{-}$, $1^{-}$, $1/2^+$, and $3/2^+$, corresponding to the pion, $\rho$ meson, nucleon, and $\Delta$ baryon, computed for a 
single lattice QCD ensemble with $a \approx 0.12~\text{fm}$ and $m_{\pi} \approx 450~\text{MeV}$~\cite{Pefkou:2021fni}.
This previous work used an ensemble with $N_f=2+1$ clover-improved dynamical quark flavors, for which the hadron GFFs receive both a quark and a gluon contribution; only the gluon one was constrained, neglecting its mixing with the quark one. The comparison of the overall normalization between those and the gluon GFFs in this work---which coincide with the total GFFs in a theory with only gluonic degrees of freedom, as investigated here---is not meaningful.
We thus rescale the results of Ref.~\cite{Pefkou:2021fni} to match each glueball GFF in the forward limit, i.e., such that $A_g(t=0)=1$ and $D_g(t=0) = D_{0^{++}}(t=0)$ for all hadrons. We can then compare the $t$-dependence of the form factors. The glueball $A(t)$ form factor decays more slowly than that of the pion, corresponding to a smaller mass radius contribution. The uncertainty of $D(t)$ is very large; however, the form factor shows a $t$-dependence more similar to that of the meson $D(t)$ form factors
than of the baryonic ones.

\section{Conclusion}

These preliminary results constitute the first time the internal structure of glueballs has been investigated 
in an $SU(3)$ lattice gauge theory,
representing a
promising first step towards understanding the internal structure of potential glueball-like hadrons in nature, and towards an analogous computation in QCD. 
The next steps
towards finalizing the calculation
include expanding the variational basis of operators to better control excited state effects, comparing results computed with different choices of EMT smearing to 
assess contamination by operator-source and operator-sink contact terms,
and extending the study to heavier glueball states with different quantum numbers.
Looking forwards, it will also be interesting to investigate whether several recent methods developments can improve the determination of these quantities.
In particular, the Lanczos analysis formalism provides more robust treatment of excited states while resolving signal-to-noise issues for both spectroscopy and matrix elements~\cite{Wagman:2024rid,Hackett:2024xnx}. 
Separately, multi-level algorithms have been shown to provide substantially improved signals in calculations of the glueball spectrum~\cite{Barca:2024fpc}, and the same technology may be applied to matrix element calculations.

\section*{Acknowledgements} 

The authors thank Julian Urban for contributions at early stages of the project. We also thank L. Barca and M. Hansen for useful discussions. This work is supported in part by the U.S.~Department of Energy, Office of Science, Office of Nuclear Physics, under grant Contract Number DE-SC0011090 and by Early Career Award DE-SC0021006, and has benefited from the QGT Topical Collaboration DE-SC0023646. PES is supported in part by Simons Foundation grant 994314 (Simons Collaboration on Confinement and QCD Strings) and by the U.S. Department of Energy SciDAC5 award DE-SC0023116.
DAP is supported from the Office of Nuclear Physics, Department of Energy, under contract DE-SC0004658.
This manuscript has been authored by Fermi Research Alliance, LLC under Contract No.~DE-AC02-07CH11359 with the U.S. Department of Energy, Office of Science, Office of High Energy Physics.
This research used resources of the National Energy Research Scientific Computing Center (NERSC), a U.S. Department of Energy Office of Science User Facility operated under Contract No.~DE-AC02-05CH11231. The authors acknowledge the MIT SuperCloud and Lincoln Laboratory Supercomputing Center for providing HPC resources that have contributed to the research results reported within this work~\cite{reuther2018interactive}. FRL acknowledges partial support by the Mauricio and Carlota Botton Fellowship.
RA is supported by the U.S. Department of Energy SciDAC5 award DE-SC0023116 and the High Energy Physics Computing Traineeship for Lattice Gauge Theory (DE-SC0024053). 

\bibliographystyle{JHEP}
\bibliography{glueballs}

\end{document}